# Quantitative Analysis of Matthew Effect and Sparsity Problem of Recommender Systems

Hao Wang
HC Research
HC Financial Service Group
Beijing, China
haow85@live.com

Zonghu Wang，Weishi Zhang
HC Research
HC Financial Service Group
Beijing, China
zhangweishi160919, wangzonghu161219@credithc.com

*Abstract*—Recommender systems have received great commercial success. Recommendation has been widely used in areas such as e-commerce, online music FM, online news portal, etc. However, several problems related to input data structure pose serious challenge to recommender system performance. Two of these problems are Matthew effect and sparsity problem. Matthew effect heavily skews recommender system output towards popular items. Data sparsity problem directly affects the coverage of recommendation result. Collaborative filtering is a simple benchmark ubiquitously adopted in the industry as the baseline for recommender system design. Understanding the underlying mechanism of collaborative filtering is crucial for further optimization. In this paper, we do a thorough quantitative analysis on Matthew Effect and sparsity problem in the particular context setting of collaborative filtering. We compare the underlying mechanism of user-based and item-based collaborative filtering and give insight to industrial recommender system builders.

## I. INTRODUCTION

Recommender system has received ubiquitous popularity in the commercial world since its introduction. E-commerce websites such as Amazon and Alibaba have been building recommender systems to recommend products to users. Online music websites like Youtube and Pandora have been using recommender systems to recommend music or videos to users. Other companies that utilize recommender systems for their commercial purposes include Goodreads , Douban , Twitter, Renren , Baidu, Tencent and Google.

Collaborative Filtering is the earliest and most widely used algorithm in the recommender system field. Researchers and industrial workers have been developing other approaches along the year such as matrix factorization methods, learning to rank approaches and deep learning paradigms. Big Chinese internet companies favor logistic regression and deep learning approaches. However, for many small and mid-sized companies, collaborative filtering is still widely adopted. Besides, collaborative filtering has been used as the most basic baseline for industrial recommender systems. Therefore, it is very important to understand the underlying mechanism of collaborative filtering.

Recommender system performance is highly dependent on input data. Several problems have been considered key challenges to recommender systems. Matthew Effect is a vital problem in recommender systems in video or news sites. To be more specific, in the context of video or news recommendation, popular items display dominating effect on recommendation performance. In particular, if collaborative filtering is the recommendation algorithm, popular items will be similar to so many other items, it will skew the recommendation results towards popular items. Another serious challenge to recommender system developers is the sparsity problem. In particular, in collaborative filtering algorithms , user ratings are predicted by computing similarities. Many times, there aren't enough similar users or items for the computed result to cover the entire data set.

Researchers and industrial workers have dealt with the aforementioned challenges for years. However, to the best of our knowledge, there has not been any quantitative analysis on how Matthew effect or sparsity problem of the input data set affects the final recommendation algorithm output. In this paper, we use collaborative filtering as our benchmark algorithm for quantitative analysis on Matthew effect and sparsity problem. We provide a thorough quantitative analysis and compare how different algorithms behave under different data contexts.

## II. RELATED WORK

Recommender system is one of the most successfully commercialized technologies in internet companies. Many companies have built dedicated teams working on recommendation products. Recommender systems could serve as standalone products or inseparable part of larger products. Baidu has been using recommender system in their Q&A site [1]. Netease adopts recommender system in their online-dating product [2]. Video recommendation is an important functionality of YouTube [3].

Collaborative Filtering has been the most widely used recommender system algorithm. It is also one of the earliest recommender system techniques. There has been tremendous literature on collaborative filtering algorithms ([4], [5], [6]) and their applications. Matthew Effect and sparsity problems have been long-term headaches for recommender system designers.

To tackle the Matthew Effect and sparsity problem, researchers and engineers have come up with many solutions. Google computes the trending categorical information to offset the Matthew Effect introduced by the most popular news articles. Bansal et. al [7] proposed a deep text recommender model which could ameliorate the sparsity problem. In his deep learning model, Wang et.al [8] utlize auxiliary information associated with items to tackle the sparsity problem.

### III. COLLABORATIVE FILTERING OVERVIEW

Recommender system is an area with long and rich history. The earliest algorithm in this area is collaborative filtering. Collaborative filtering could be categorized into user-based collaborative filtering and item-based collaborative filtering. User-based collaborative filtering computes the similarity score between users and weight the items of most similar users to compute the final rating for a product by the user. Item-based collaborative filtering computes similarity score between items first, and uses a similar approach to compute user's ratings.

Matthew Effect is a common and well-known phenomenon existing in many real world context settings. For example, the degrees of vertices in a social network generally obeys the power law distribution. One of the most simplified probability model used to quantify Matthew Effect is Zipf's Law. Zipf's Law captures the Matthew Effect using the following probability mass function :

$$f(k; s, N) = \frac{1/k^s}{\sum_{n=1}^{N}(1/n^s)}$$

, where N is the total number of elements in consideration , k is the rank of a given item , and s is a parameter.

In this paper we use the following metrics to quantify the Matthew Effect and sparsity problem of collaborative filtering techniques:

1. **Matthew Effect** : To evaluate the Matthew Effect of different approaches, we analyze the expected similarity score between users (user-based collaborative filtering) and the expected similarity score between items (item-based collaborative filtering). To be more specific, we investigate how popularity effects the similarity score between the selected user / item and other users / items. The degree to which such similarity score exhibits disproportional larger value for more popular user / item shows the severity of the Matthew effect, which we demonstrate how to quantify analytically.

2. **Sparsity Problem:** To evaluate the sparsity problem of different approaches, we use the expected users (user-based collaborative filtering) and expected items (item-based collaborative filtering) involved in the similarity score computation. The more users / items get involved in computation , the data set gets less sparse. We compute the metrics using combinatorics and compare the sparsity effect in user-based and item-based collaborative filtering approaches.

### IV. MATTHEW EFFECT IN USER BASED COLLABORATIVE FILTERING

Let's consider the context of recommender system as follows : We recommend videos to users. Users could click on the videos to view them , we would like to build a recommender system based on the user click log. In general , user click distribution obeys long tail distribution. To simplify our model, we assume the user click distribution is Zipf's Law Distribution. That is, if the number of clicks on the most popular video is N, then the number of clicks on the i-th most popular video is N/i . In addition, we assume users' click behavior is independent with each other.

Consider two users A and B, user A viewed $N_A$ videos, and user B viewed $N_B$ videos. The set of videos watched by user A is $I_A$ , and the set of videos watched by user B is $I_B$. The number of videos in total is M. The probability of the i-th most popular video appearing in a user's click log is :

$$\frac{\frac{1}{i}}{\sum_{j=1}^{M}\frac{1}{j}}$$

The probability that this video is clicked by both user A and user B is:

$$\left[\frac{\frac{1}{i}}{\sum_{j=1}^{M}\frac{1}{j}}\right]^2$$

In our video recommendation context setting , the user-based collaborative filtering algorithm follows the following formula :

$$R(i, j) = \frac{\sum_{k=1}^{N} sim(i,k) \times I(sim(i,j) > 0)}{\sum_{j=1}^{N} I(sim(i,j) > 0)}$$

In other words, R(i, j) is the average of user i's similarity with other users. The crucial step of predicting a user's likeness of an item is the computation of the user's similarity with another user. We use the Jaccard distance as the similarity metric between two users, that is, the similarity score between two users is the number of videos clicked by both of them divided by the total number of videos clicked by either of them.

If user A and user B share only 1 video, their expected similarity score is:

$$\sum_{i=1}^{M} \left( \frac{\frac{1}{i}}{\sum_{j=1}^{M} \frac{1}{j}} \right)^2 \frac{1}{|I_A \cup I_B|}$$

If user A and user B share 2 videos, their expected similarity score is:

$$\sum_{i=1}^{M-1} \sum_{k=i+1}^{M} \left( \frac{\frac{1}{i}}{\sum_{j=1}^{M} \frac{1}{j}} \right)^2 \left( \frac{\frac{1}{k}}{\sum_{j=1}^{M} \frac{1}{j}} \right)^2 \frac{2}{|I_A \cup I_B|}$$

In general, the expected similarity score between user A and user B is:

$$\sum_{t=1}^{\min(N_A,N_B)} \left[ \sum_{i_1=1}^{M} \sum_{i_2=i_1+1}^{M} \cdots \sum_{i_t=i_{t-1}+1}^{M} \left( \frac{\frac{1}{i_t}}{\sum_{j=1}^{M} \frac{1}{j}} \right)^2 \times \cdots \times \left( \frac{\frac{1}{i_t}}{\sum_{j=1}^{M} \frac{1}{j}} \right)^2 \right] \times \frac{t}{|I_A \cup I_B|}$$

## V. MATTHEW EFFECT IN ITEM BASED COLLABORATIVE FILTERING

Let's consider two videos: video A is clicked by 1/m of the users, video B is clicked by 1/n of users. To simplify our statistical model, we assume the event video A is clicked is independent of the event video B is clicked. The probability video A and video B are clicked by the same user is then 1/(m*n). Assume there are W users in total, then the event that a user clicked two videos obeys the Bernoulli distribution. On average, there are W/(m*n) users who clicked both videos. We also know that, on average, W/m users clicked video A and W/n users clicked video B.

If we use L1-norm cosine distance to compute the similarity score between video A and video B, then we have:

$$Sim(A,B) = \frac{\frac{W}{m*n}}{\frac{W}{m} * \frac{W}{n}} = \frac{1}{W}$$

If we use L2-norm cosine distance to compute the similarity score between video A and video B, then we have:

$$Sim(A,B) = \frac{\frac{W}{m*n}}{\sqrt{\left(\frac{W}{m}\right)\left(\frac{W}{n}\right)}} = \frac{1}{\sqrt{m*n}}$$

From the similarity score, it is obvious L1-norm cosine is not affected by the Matthew effect of video distribution, since sim(A,B) is not dependent on m or n. However, when using L2-norm, sim(A,B) is dependent on m and n. In other words, sim(A,B) is dependent on the number of users clicking either video. The more clicks a video receives, the higher the similarity score will be. Matthew effect greatly influences the final recommendation results.

## VI. SPARSITY IN USER BASED COLLABORATIVE FILTERING

In the user collaborative filtering context setting, assume there are N users and M videos in total. The probability that a user clicks the i-th most popular video is:

$$\frac{\frac{1}{i}}{\sum_{j=1}^{M} \frac{1}{j}}$$

The expected number of other users that click on the i-th most popular video is:

$$(N-1) \times \frac{\frac{1}{i}}{\sum_{j=1}^{M} \frac{1}{j}}$$

The expected number of users involved in the similarity computation with a given user is :

$$\sum_{i=1}^{N} (N-1) \times \frac{\frac{1}{i}}{\sum_{j=1}^{M} \frac{1}{j}}$$

From the analytical result, we know that the sparsity problem of the user collaborative filtering could be modeled using Zipf's law distribution. The user based collaborative filtering is highly susceptible to the skewness of the input data structure.

## VII. SPARSITY IN ITEM BASED COLLABORATIVE FILTERING

To evaluate the sparsity effect of collaborative filtering systems, we compute the expected number of videos that is involved in the similarity computation with a given video.

Assume click behavior among users are independent, also clicks of a given user are independent with each other. The probability that a user clicks both the i-th most popular and the j-th popular videos is :

$$P_u(i, j) = \frac{1}{i} \times \frac{1}{j}$$

The expected number of videos involved in the similarity computation of the i-th most popular video is :

$$\sum_{j=1, j \neq i}^{N} \frac{1}{i} \times \frac{1}{j}$$

From the formula , it is apparent the distribution of the similarity computation item number of videos obeys Zipf's Law. In other words, let the number of videos involved in similarity computation of the i-th most popular video be N(i), then we have :

$$\frac{N(i)}{N(j)} = \frac{\frac{1}{i}}{\frac{1}{j}} = \frac{j}{i}$$

## VIII. EXPERIMENT

We explore the Matthew Effect and sparsity problem of collaborative filtering algorithms using the Lastfm data set. The Lastfm dataset contains a list of users and the artists of songs that they listen to. The number of users is 1892 and the number of artists is 17632.

We use the user-based collaborative filtering context and compute the average similarity score between user pairs of different ranks and plot the result in Fig. 1

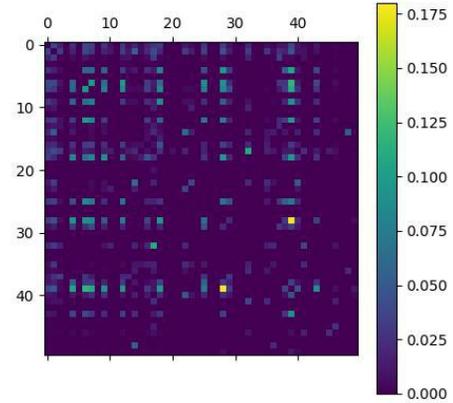

Fig.1 Similarity scores between Lastfm user pairs of different ranks using user-based collaborative filtering

From Fig.1 it is pretty obvious how Matthew Effect of the input data structure leads to skewness in similarity score of users of different ranks. Higher similarity scores concentrate in the upper left corner of the similarity score matrix.

Matthew Effect often leads to unwanted results user-based collaborative filtering. The implication of Fig. 1 is that users with more items tend to be similar with users with more items. This means users with diverse interest will be recommended with more diverse items, while users with less diverse interest will hardly get items of other categories. This is contradictory to one of recommender systems' crucial missions: Helping users find more interesting items that are otherwise hard to encounter. In addition, if the diversity of the recommended items is too high , it's very difficult to capture the real taste of the users, especially those who already exhibited diversified tastes.

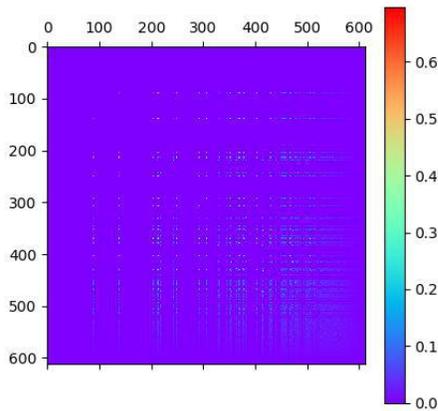

Fig 2. Similarity scores between Lastfm item pairs of different ranks using item-based collaborative filtering with Jaccard distance

Fig 2. demonstrates the Matthew Effect of similarity scores in item-based collaborative filtering as a consequence of skewness in the input data structure. The orange sparkles at the upper left corner and the spread light blue points at the lower right corner illustrate the algorithmic structure of the data set : Popular items are seldom similar to each other, but when they do , they have a rather high similarity score .

We also compute the Matthew Effect metrics in collaborative filtering context settings. Fig. 3 shows the number of users involved in computation of similarity scores of users of different ranks. The distribution of data points exhibit power-law distribution properties. According to our quantified metrics, the distribution of these points should follow Zipf's Law distribution if the input data follows Zipf's Law distribution. The real world dataset proves the correctness of our formulas.

Fig 4. shows the Matthew Effect metrics for item-based collaborative filtering. The x-axis is the items of different ranks while the y-axis is the number of items involved in computation of similarity scores. The distribution of data also follows a distribution similar to Zipf's Law. This once again proves the formulas of our quantified results.

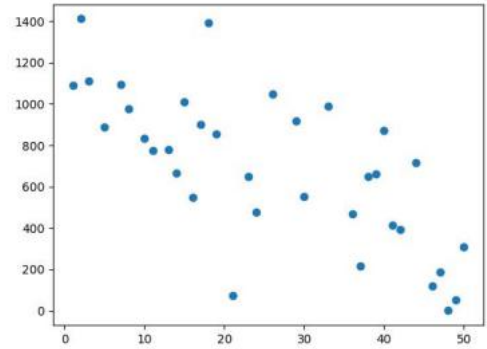

Fig. 3 Number of users involved in similarity score computation of users of different ranks in user based collaborative filtering algorithm

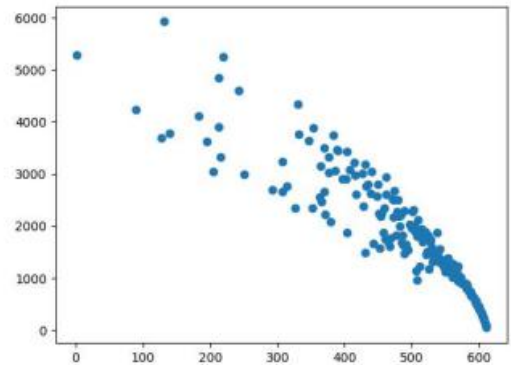

Fig. 4 Number of items involved in similarity score computation of items of different ranks in item based collaborative filtering algorithm

## 7 CONCLUSION

Collaborative filtering is the most widely used recommender system technique. Its simplicity is very helpful in analyzing the data structure for more complicated algorithms. Understanding how Matthew Effect and sparsity problem affects the collaborative filtering approaches helps understand other algorithms. In this paper, we proposed metrics used to quantify Matthew Effect and the sparsity problem. We computed the metrics analytically for user-based collaborative filtering and item-based collaborative filtering. We provided a theoretic foundation for the analysis of Matthew Effect and sparsity problem of recommender system approaches.

In future research, we would like to explore quantitative analysis of other recommender system models such as matrix factorization , learning to rank and deep learning.

Besides collaborative filtering, we would also like to find out how input data structure affects these other algorithms' output. We hope our research work could help industrial researchers and engineers design better algorithms and systems.